# Critical Remarks on Landauer's principle of erasure–dissipation

Including notes on Maxwell demons and Szilard engines

L.B. Kish
Department of Electrical Engineering
Texas A&M University
College Station, TX 77843-3128, USA

C.G. Granqvist
Department of Engineering Sciences
The Ångström Laboratory, Uppsala University,
P.O. Box 534, SE-75121 Uppsala, Sweden

S.P Khatri
Department of Electrical Engineering
Texas A&M University
College Station, TX 77843-3128, USA

J.M. Smulko
Department of Metrology and Optoelectronics
Faculty of Electronics, Telecommunications and Informatics
Gdansk University of Technology
Narutowicza 11/12, 80-233 Gdansk, Poland

*Abstract*—We briefly address Landauer's Principle and some related issues in thermal demons. We show that an error-free Turing computer works in the zero-entropy limit, which proves Landauer's derivation incorrect. To have a physical logic gate, memory or information-engine, a few essential components necessary for the operation of these devices are often neglected, such as various aspects of control, damping and the fluctuation–dissipation theorem. We also point out that bit erasure is typically not needed or used for the functioning of computers or engines (except for secure erasure).

## I. Introduction

Szilard [1] (in 1929, in an incorrect way) and Brillouin [2] (in 1953, in the correct way) concluded that the minimum energy dissipation $H_1$ due to changing a bit of information in a system at absolute temperature $T$ is given as

$$H_1 \geq kT \ln(2), \qquad (1)$$

where $k$ is the Boltzmann constant. Later Brillouin [3], Kish [4] and Alicki [5], for *arbitrary bit flips (writing or erasure)*, refined this equation, in independent ways, to read

$$H_1 \approx kT \ln\left(\frac{1}{\varepsilon}\right), \qquad (2)$$

where $\varepsilon \leq 0.5$ is the bit-error probability of the operation. Equation (1) leads to Eq. (2) in the limit $\varepsilon = 0.5$, where the efficiency of the operation is zero. Equations (1) and (2) are valid when the measurement time window is short compared to the correlation time $\tau$ of thermal fluctuations and, for longer time windows $t_w \geq \tau$, Kish and Granqvist [6,7] gave a correction in the low-error limit according to

$$H_1 \approx kT \left[ \ln\left(\frac{1}{\varepsilon}\right) + \ln\left(\frac{t_w}{\tau}\right) \right]. \qquad (3)$$

In contrast to the above treatments, Landauer [8] and his principle claims that *only erasure is dissipative*, i.e., the $1 \rightarrow 0$ bit flip, because that operation involves shrinking the space-state from two possible bit-values (0,1) into a determined single bit value (0) and a concomitant reduction of the entropy in the memory, which must be compensated by a corresponding entropy production (heat) in the rest of the system, given by Eq. (1). In Section III, we will point out why this argumentation is flawed and correct other errors too.

## II. The Standoff

Porod, Ferry, *et al*. [9–11] as well as Norton [12,13] have extensively analyzed and refuted Landauer's principle; see Sec. 3 below. Later Gyftopoulos and Spakovsky [14] and Kish [6,7,15] also refuted it. Currently there is no conversation between members of the *pro* side, who have published a vast body of papers in leading journals and magazines on Landauer's principle and related issues, and members of the *con* side, which are fewer in number but voice strong concrete objections.

There are even experimental papers that claim to "prove" Landauer's principle, for example [16]. Unfortunately they are flawed, see III/C (*Controlling the barrier*) for illustration.

In the rest of this paper, we address a few important aspects of the criticism.

## III. MEMORIES AND GATES: FUNDAMENTAL ERRORS

### A. The core flaw: Misunderstanding entropy.

The core argument leading to the Landauer Principle is summarized by Bennett [17] as follows:

*Suppose a memory register of n bits is cleared; in other words, suppose the value in each location is set at zero, regardless of the previous value. Before the operation the register as a whole could have been in any of $2^n$ states. After the operation the register can be in only one state. The operation has therefore compressed many logical states into one, much as a piston might compress a gas.* [17]

If the above statement were correct, such space state reduction would reduce the entropy in the memory and, due to the Second Law of Thermodynamics, that would necessitate a compensating entropy production out of the memory indeed implying corresponding heat dissipation. However the argumentation is invalid. Porod *et al*. [9,10], Porod [10] and Norton [12,13] point out that the logic state in a computer is different from the system state in a thermodynamic ensemble. Let us discuss a single-bit example (Fig. 1). It is similar to Landauer's example [8], but we reach a totally different conclusion.

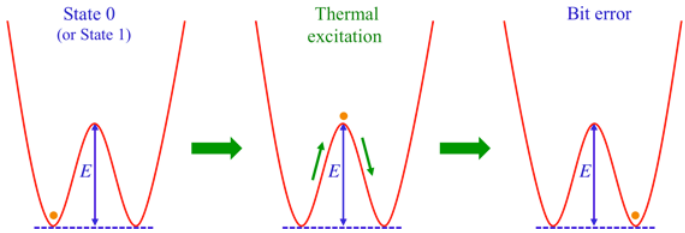

Fig. 1. Model example for a non-volatile memory: particle in a double potential well (simulating, for example, a magnetic memory bit). Bit value 0: particle in the left well; bit value 1: particle in the right well (without loss of generality). When thermal fluctuations generate a spontaneous bit flip, a bit error emerges. $H_1$ in Eqs. (2) and (3) is implied by moving the particle over the energy barrier $E$ and is equal to it.

The entropy $S$ of a two-state system is given as

$$S = -k(p_0 \ln p_0 + p_1 \ln p_1), \quad (4)$$

where $p_0$ and $p_1$ are the probability of being in the state with bit values 0 and 1, respectively. Landauer [8] and Bennett [17,18] implicitly assume that, before erasure, the probabilities are

$$p_{0b} = p_{1b} = 0.5, \quad (5)$$

while, after erasure,

$$p_{0a} = 1 \quad \text{and} \quad p_{1a} = 0. \quad (6)$$

Thus they envision a reduction of state-space by a factor of two and a corresponding entropy reduction of

$$\Delta S = -k[0.5\ln(0.5) + 0.5\ln(0.5)] = k\ln(2), \quad (7)$$

because the entropy is zero after the erasure.

The fundamental error lies in Eq. (5), which would be valid if we waited for such a long time that the information in the memory got fully destroyed by thermal fluctuations. Thus the memory can contain the incorrect or correct bit-value with the same probability. However, this is the situation we must avoid by all means in deterministic Turing machine computers, and we must keep the bit-error probabilities at virtually zero. Thus the corrected Eq. (5) is

$$\begin{array}{c} p_{0b} = 0 \quad \text{and} \quad p_{1b} = 1 \\ \text{or} \\ p_{0b} = 1 \quad \text{and} \quad p_{1b} = 0 \end{array}. \quad (7)$$

The corresponding correct entropy of the system before and after erasure is zero, *i.e.*,

$$S_b = S_a = 0. \quad (8)$$

Consequently the entropy reduction is zero! Obviously, writing the bit value 1 into the memory in 0 state results in the same entropies.

In conclusion, erasure does not have any special role in physical informatics as a consequence of the deterministic nature of Turing machines. This conclusion remains the same when we account for thermal noise, because both the 0 and the 1 states will be less deterministic to the same degree (see Fig. 1). In fact during both writing and erasure, the dissipation limits given by Eq. (3) hold under general (small-error) conditions.

Thus the question emerges: where does then the heat generation come from? This question is answered in *C* below.

### B. Logic reversibility does not imply physical reversibility

A related issue is the postulation of reversible computation. In such a computer [17,18] the logic sequence of operations would be logically reversible, with the exception of a (perhaps smaller) memory that records the information necessary for the reversal. By postulating that logical reversibility implies the possibility of physical reversibility, Bennett and Landauer introduced reversible computers wherein only that smaller memory would dissipate heat and do so only during erasure – if Landauer's Principle were valid.

However, as Brillouin [2,3], Porod *et al*. [9,10], Porod [10], Norton [12,13], Alicki [5] and Kish *et al*. [4,6,7,15] point out, any bit-flip is dissipative. In fact, Eq. (3) provides the lower limit of heat dissipation during a single bit-flip.

*C. Incomplete system model: Neglecting noise and damping*

In a vast body of literature (not cited here), the source of error is the understanding of what functioning systems—such as logic gates, memories or demon—entail. Essential components are neglected, and the analysis and conclusions are made on systems that could not possibly function or are simply unphysical. The clarification of these issues shows the fundamental sources of dissipation, too. We mention here only three examples: damping, noise, and controlling the potential barriers.

*Damping:* Logic gates, memories, switches, *etc.*, require damping to annihilate their excess energy after state-change. Figure 1 serves as an example: when the state is changed by deterministically pushing the particle over the barrier and into the other well, its accumulated kinetic and potential energy must be dissipated since otherwise this energy will bounce back and forth between the two wells.

*Noise*: Authors often neglect noises by saying they are discussing "idealized" systems. The designation "idealized" is incorrect in this case; it could be a "zero-temperature" system or a "damping-free" system, but the first one is unphysical because the authors assume room temperature, and the second one prohibits function as argued above. The implication of finite temperature and damping is noise in accordance with the Fluctuation–Dissipation Theorem. By supposing zero noise, the system will be unphysical under the assumed conditions of study.

*Controlling the barrier*: The other way of changing the particle location (Fig. 1) is a proper sequence of controlled barrier shape change, which ends the particle in the other well and flips the bit value. This is actually happening in CMOS technology (an "off" switch is a large potential barrier while an "on" switch is a zero barrier) and Landauer [8] and an experimental "proof" [16] also used this method. While Landauer [8] and the experimental paper [16] fully neglected the energy dissipation requirements of barrier control. On the other hand, in today's CMOS chip technology, the barrier control (running the switches) yields the dominant energy dissipation. More on the energy requirement of control is shown in the next section and papers [6,7].

IV. DEMONS: NEGLECTED DISSIPATION OF CONTROL [15]

Surprisingly, Landauer's Theorem has been connected with thermodynamic demons such as Maxwell demons, Szilard engines and Quantum demons. We are unable to cite all of the vast literature on this topic; instead we refer to a few examples [18,19].

Equations (1) and (2) make it obvious [6,7] that whenever a single switching or decision operation takes place during an engine cycle, then an energy exceeding $kT\ln(2)$ is dissipated during each of these operations. This is extremely important, because the energy the demon is supposed to produce is exactly this one (Szilard engine), or at least has a similar value (Maxwell demon). Scrutinizing a paper by Bennett [18], all or almost all operations of a single demon cycle involve such single-bit operations, which means that the control steps in themselves generate many times greater heat than the one given by the $kT\ln(2)$ limit [6,7]. Words such as "insert", "observe", "if", "go to", "attach", (letting to) "expand", "remove", "transform", *etc.*, signify such minimum single-bit control operations in the mentioned paper [18]. Specific operations are exemplified below from Bennetts work [18], where L/R means left/right and M1 … L3 are designations used by Bennett [16] (we added Italic typeface):

M1. *Insert* partition [L]
M2. *Observe* the particle's chamber [L] or [R]
M3. *If* memory bit = R, *go to* R1 [R]
M4. *If* memory bit = L, *go to* L1 [L]
R1. *Attach* pulleys so right chamber can expand [R]
R2. *Expand*, doing isothermal work $W$ [R]
R3. *Remove* pulleys [R]
R4. *Transform* known memory bit from R to L [L]
R5. *Go to* M1 [L]
L1. *Attach* pulleys so left chamber can expand [L]
L2. *Expand*, doing isothermal work $W$ [L]
L3. *Remove* pulleys [L] L4. Go to M1 [L]

Even Szilard [1] missed the crucial aspect of control energy in his famous paper, because he used a two-stage lever to control a gear. The control of this gear consumes all of the energy the Szilard engine was supposed to produce [6,7]. Thus the Szilard-engine-puzzle about the Second Law and intelligent being is non-existent from the beginning. In hindsight, Szilard's error might be considered fortunate because it led to an expansion of the research field. In the case of Szilard's engine, it is possible to fix the mentioned flaw [6,7], but the question as to the role of an intelligent being remains obsolete.

We note that Renner [19] and others claim that control energies can be saved by using a large number of parallel demons mechanically coupled to each other. This belief is incorrect since demons are fundamentally random so that they require independent control units and related energy dissipation. Conservation of control energy can be done in periodic engines, where this effect can be executed both in space (parallel-coupled engines) and time (resonators with high Q-factor) [20], but it is not possible for demons.

V. ERASURE DISSIPATION IN PRACTICAL COMPUTING [15]

Finally, we address the problem of non-secure bit erasure in large memories in computers. Do computers execute erasure when they discard information? The answer is usually "no" since, in practice, they do not reset the memory bits but just change the address of the boundary of the free part of memory as illustrated in Fig. 2. The number of bits in the address scales as $\log_2(N)$, where $N$ is the size of the whole memory, and hence (in accordance with Eq. 3) the energy dissipation is of the order of

$$H_N \approx kT\left[\ln\left(\frac{1}{\varepsilon}\right)+\ln\left(\frac{t_w}{\tau}\right)\right]\log_2(N). \qquad (9)$$

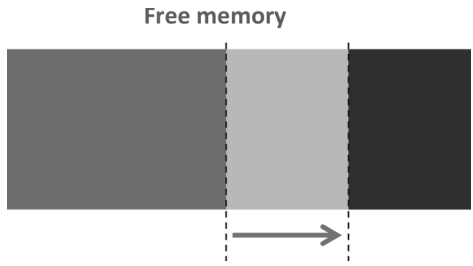

Fig. 2. Simplified one-dimensional illustration of discarding information in an idealized computer memory. Gray tones signify space free for writing, and black denotes occupied space. The address of the boundary of the free memory is moved along the arrow to discard information and increase the free-memory part.

At fixed error rate and observation time window, the energy dissipation of erasure scales as

$$H_N \propto \log_2(N), \qquad (10)$$

which is a much more optimal situation than for the Landauer limit of $kT\ln(2)$ energy dissipation per bit. Thus, even if Landauer's Principle were valid, it would still be of limited practical importance.

ACKNOWLEDGMENT

LBK is grateful for extensive discussions with John Norton, Wolfgang Porod, Dave Ferry and Xavier Oriols.